\documentclass[11pt,a4paper]{article}
\usepackage[english]{babel}
\usepackage{amsfonts,amsbsy,bm,euscript,mathrsfs}
\usepackage{amssymb,stmaryrd,faktor,slashed}
\usepackage[x11names]{xcolor}
\usepackage[tbtags]{amsmath}
\usepackage[nosort]{cite}

\numberwithin{equation}{section}

\makeatletter
\renewcommand\section{\@startsection {section}{1}{\z@}
{-3.5ex \@plus -1ex \@minus -.2ex}
{2.3ex \@plus.2ex}
{\normalfont\Large\bfseries}}
\renewcommand\subsection{\@startsection{subsection}{2}{\z@}
{-3.25ex\@plus -1ex \@minus -.2ex}
{1.5ex \@plus.2ex}
{\normalfont\large\bfseries}}
\makeatother

\def\ads{{\rm AdS}_5\times {\rm S}^5}

\begin{document}

\setcounter{equation}{0}
\setcounter{footnote}{0}
\setcounter{section}{0}

\thispagestyle{empty}

\begin{flushright}
\texttt{
HU-EP-18/12
\\ HU-MATH-18/04
}
\end{flushright}

\begin{center}

\vspace{1.5truecm}

{\LARGE \bf On Yang-Baxter models, twist operators, and boundary conditions}

\vspace{1.5truecm}

{Stijn J. van Tongeren}

\vspace{1.0truecm}

{\em Institut f\"ur Mathematik und Institut f\"ur Physik, Humboldt-Universit\"at zu Berlin, \\ IRIS Geb\"aude, Zum Grossen Windkanal 6, 12489 Berlin, Germany}

\vspace{1.0truecm}

{{\tt svantongeren@physik.hu-berlin.de}}

\vspace{1.0truecm}
\end{center}

\begin{abstract}
We discuss homogeneous Yang-Baxter deformations of integrable sigma models in terms of twist operators. We show that the twist operators behave as the classical analogue of a Drinfeld twist, for all abelian and almost abelian deformations. We also use twist operators to rederive the well-known interpretation of TsT transformations -- equivalent to abelian deformations -- in terms of twisted boundary conditions. We discuss complications in extending this boundary condition picture to non-abelian deformations.
\end{abstract}

\newpage

\setcounter{equation}{0}
\setcounter{footnote}{0}
\setcounter{section}{0}

\tableofcontents

\section{Introduction}

The integrability of the $\ads$ superstring allows for it to be solved exactly, at finite coupling, by Bethe ansatz type methods \cite{Arutyunov:2009ga,Beisert:2010jr,Bombardelli:2016rwb}. The same methods can be extended to certain integrable deformations of this string, namely the $\beta$ deformation \cite{Lunin:2005jy} and the $\eta$ deformation \cite{Delduc:2013qra}, see e.g. \cite{Klabbers:2017vtw} and references therein. These models are examples of Yang-Baxter deformations \cite{Klimcik:2002zj,Klimcik:2008eq,Delduc:2013fga} of the $\ads$ string. In this setting, the $\eta$ deformation of \cite{Delduc:2013qra} is an inhomogeneous deformation, while the $\beta$ deformation is a simple homogeneous deformation \cite{Kawaguchi:2014qwa,Matsumoto:2014nra}. While the inhomogeneous $\eta$ deformation is essentially unique, there are many different homogenous deformations. It is an open question how to describe general homogeneous deformed models at the quantum level. The technology built for the undeformed model can, in essence, be adapted to the $\beta$ deformation, because the deformed model can be mapped back to the undeformed model with non-periodic (twisted) boundary conditions \cite{Frolov:2005dj,Alday:2005ww}, and these boundary conditions are compatible with the Bethe ansatz. Here we discuss a setup that reproduces these results, and that can be extended to generic homogeneous models. We will discuss how generic homogeneous models differ significantly from the $\beta$ deformation, leading to interesting and challenging open problems.

The analysis of \cite{Frolov:2005dj} is built on the interpretation of the $\beta$ deformation as a T duality-shift-T duality (TsT) transformation. The resulting duality relations between the coordinates, which can be interpreted as a non-local field redefinition, lead to the relation between the deformed model and the undeformed one with twisted boundary conditions. This analysis is not limited to the $\beta$ deformation, and applies to an arbitrary sequence of commuting TsT transformations \cite{Alday:2005ww}. In Yang-Baxter terminology, such transformations are equivalent to abelian deformations \cite{Osten:2016dvf}. Generic homogeneous deformations are equivalent to non-abelian T duality transformations \cite{Hoare:2016wsk,Borsato:2016pas}, see e.g. \cite{Lust:2018jsx} and references therein for recent developments.

The main goal of this note is to set up a clean framework to start to systematically generalize aspects of the abelian boundary condition picture to non-abelian cases. To do so, we work in the twist operator picture of \cite{Matsumoto:2015jja} for homogeneous models, see also \cite{Vicedo:2015pna}. In this picture, arbitrary homogeneous deformations can be formally rephrased as non-local redefinitions of the undeformed group valued fields, expressed via a twist operator. We will discuss how this twist operator can be expressed in terms of the $R$ operator defining the deformation, and a matrix current that contains the conserved Noether currents of the deformed model. In the abelian case we can use this to see that the twist operator manifestly behaves as the classical analogue of a Drinfeld twist. Moreover, the twist operator naturally encodes the known boundary condition picture for TsT transformations -- the type of relation we would like to generalize. For non-abelian homogeneous deformations, however, the interpretation of the twist operator is more complicated. Due to the non-abelian nature of the twist operator it is in general not possible to pick sigma model fields that result in diagonalized twisted boundary conditions expressed in terms of conserved charges. As an illustrative example we consider almost abelian deformations, where we can also see that the twist operator behaves as the analogue of a Drinfeld twist. These explicit links between twist operators and Drinfeld twists moreover provide further support for the argumentation of \cite{vanTongeren:2015uha}.

In the next section we give a brief overview of homogeneous Yang-Baxter deformations of principal chiral models, and discuss how to rephrase them via twist operators. Next we briefly recall the link between Drinfeld twists and $r$ matrices, and recall how the twist operator affects the monodromy matrix. Then in section \ref{sec:globalsymmetries} we discuss the Noether currents of deformed models, and their relation to the twist operator. In section \ref{sec:abeliandeformations} we cover abelian deformations, followed by non-abelian ones in section \ref{sec:abeliandeformations}. We conclude by summarizing open challenges, at both the classical and the quantum level. In the main text we focus on deformations of principal chiral models, providing a corresponding summary for deformations of symmetric space models in appendix \ref{app:symmetricspacediscussion}.

\section{Yang-Baxter models}

Consider a principal chiral model based on a (semi-)simple Lie group $G$ with Lie algebra $\mathfrak{g}$. The Yang-Baxter deformation of this model \cite{Klimcik:2002zj,Klimcik:2008eq} is based on an antisymmetric operator $R:\mathfrak{g}\rightarrow\mathfrak{g}$ which solves the classical Yang-Baxter equation, guaranteeing integrability of the resulting model. We will consider only homogeneous Yang-Baxter deformations \cite{Matsumoto:2015jja}, where $R$ satisfies
\begin{equation}
\label{eq:operatorCYBE}
[R(x),R(y)] - R([R(x),y]+[x,R(y)])=0, \quad \forall x,y \in \mathfrak{g},
\end{equation}
the homogeneous classical Yang-Baxter equation. Antisymmetry means $\mbox{Tr}(R(x)y)=-\mbox{Tr}(xR(y))$. Given $g\in G$ we construct a current $A = -g^{-1} dg$, and use it to define the Lagrangian
\begin{equation}
\mathcal{L} = P_-^{\alpha\beta}\mbox{Tr}\left(A_\alpha \frac{1}{1-\eta R_g}(A_\beta)\right),
\end{equation}
where $R_g = \mbox{Ad}_{g}^{-1} \circ R \circ \mbox{Ad}_g$, indices $\alpha,\beta,\ldots$ take values corresponding to the worldsheet coordinates $\tau$ and $\sigma$, and $P_\pm^{\alpha\beta} = \tfrac{1}{2}(\gamma^{\alpha\beta} \pm \epsilon^{\alpha \beta})$ with $\gamma^{\alpha\beta} = \sqrt{-h} h^{\alpha\beta}$, $h$ the worldsheet metric, and $\epsilon^{\tau\sigma}=-\epsilon^{\sigma\tau}=1$. The $P_\pm$ project one-forms onto their Hodge (anti-)self-dual components, i.e. $\star X_\pm = \pm X_\pm$, where $X_\pm^\alpha \equiv P_\pm^{\alpha \beta} X_\beta$.\footnote{The projection operators come with various other useful identities such as $P^{\beta\alpha}_\pm = P^{\alpha\beta}_\mp$, $P_\pm^{\alpha \delta}\gamma_{\delta\zeta}P_\pm^{\zeta \beta} = P_\pm^{\alpha\beta}$, $P_\pm^{\alpha \delta}\gamma_{\delta\zeta}P_\mp^{\zeta \beta} = 0$, and the related $X_\alpha Y^{\alpha}_\pm = X_{\mp\alpha}Y^{\alpha}_\pm$. }

We get a simple description of this model in terms of the current
\begin{equation}
I =  \frac{1}{1+\eta R_g}(A_+) + \frac{1}{1-\eta R_g}(A_-).
\end{equation}
Conversely,
\begin{equation}
\label{eq:AviaI}
A = \left(1+ \eta R_g\, \star \right) I.
\end{equation}
For future reference, we also consider the left current $A^l = \mbox{Ad}_g(A)$, with accompanying
\begin{equation}
I^l = \mbox{Ad}_g(I) = \frac{1}{1+\eta R}(A^l_+) + \frac{1}{1-\eta R}(A^l_-).
\end{equation}
The equations of motion of the model are
\begin{equation}
\mathcal{E} = \partial_\alpha I^\alpha = 0,
\end{equation}
or in terms of left currents
\begin{equation}
\label{eq:lefteom}
\mathcal{E}^l = \partial_\alpha I^{l\alpha} - \eta \epsilon^{\alpha \beta}[R(I^{l}_\alpha),I^{l}_\beta] = 0,
\end{equation}
demonstrating the relative elegance of working in the right formulation for a left deformation.

The flatness of $A$ expressed in terms of $I$, using the homogeneous classical Yang-Baxter equation, becomes
\begin{equation}
\label{eq:Iflat}
\epsilon^{\alpha\beta}\left(\partial_\alpha I_\beta - \partial_\beta I_\alpha - [I_\alpha,I_\beta]\right)= - 2 \eta R_g(\mathcal{E}).
\end{equation}
We see that, on shell, $I$ is flat. We can take the associated Lax connection to be
\begin{equation}
L(z) =\frac{1}{1+z} I_+ + \frac{1}{1- z} I_-,
\end{equation}
where $z$ is the spectral parameter.

\subsection*{Twist operator}

On shell, everything about the deformed models looks exactly like the undeformed model -- just replace $A$ by $I$. We can try to translate this relation back to the group level, inspired by similar discussions in the symmetric space case \cite{Matsumoto:2015jja}. As equation \eqref{eq:Iflat} shows, on shell $I$ is flat, meaning that then we can parametrize it as a standard right current, $I = -\tilde{g}^{-1} d \tilde{g}$. By definition $I$ and $A$ are then related by the gauge transformation
\begin{equation}
\label{eq:gaugerelationIandA}
I = f A f^{-1} + df f^{-1}, f =\tilde{g}^{-1} g.
\end{equation}
If we now define the twist operator $\mathcal{F}$ as an explicit relation between $\tilde{g}$ and $g$
\begin{equation}
\label{eq:gisFgtilde}
g= \mathcal{F} \tilde{g},
\end{equation}
we find that it must solve the fundamental linear problem
\begin{equation}
d \mathcal{F} = (I^l - A^l) \mathcal{F}.
\end{equation}
Using the left analogue of equation \eqref{eq:AviaI} this means
\begin{equation}
\label{eq:dFisRF}
d\mathcal{F} = - \eta\,R(\star I^{l}) \mathcal{F}.
\end{equation}
We can formally solve for $\mathcal{F}$ in monodromy form as
\begin{equation}
\mathcal{F}(\tau,\sigma) = \mathcal{P}\left\{ \exp \left(\eta \int_0^\sigma R(I^{l\tau}) d\sigma^\prime\right)\right\} \, \mathcal{F}(\tau,0).
\end{equation}
This gives a non-local expression for $\tilde{g}$. Note that the definitions of $A$ and $I$ as right currents, together with equations \eqref{eq:gisFgtilde} and \eqref{eq:dFisRF}, imply the original definition of $I$ above.

Formally, undeformed quantities become their deformed counterparts upon replacing $g$ by $\tilde{g}$. We will use accompanying notation, where we use a tilde to indicate replacing $g$ by $\tilde{g}$ in an undeformed expression, e.g. $\tilde{A} = -\tilde{g}^{-1} d \tilde{g} = I$.

\section{Drinfeld twisted symmetry and $r$ matrices}

\label{sec:Drinfeldtwistedsymmetryandrmatrices}

At the level of the symmetry algebra, inhomogeneous Yang-Baxter deformations are known to correspond to trigonometric $q$ deformations \cite{Delduc:2013fga,Delduc:2017brb}, in line with general expectations based on deformation quantization theory. By analogy, as argued in more detail \cite{vanTongeren:2015uha}, we expect homogeneous deformed models to have Drinfeld twisted symmetry. This idea is also supported by analysis of specific models, see e.g. \cite{Ahn:2010ws,Kawaguchi:2013lba}. In this picture, it would be natural if the above twist operator is (closely related to) the classical analogue of a Drinfeld twist.

Drinfeld twists are used to deform Hopf algebras. Relatedly, they appear in twisted quantum integrable models, where a Drinfeld twist $F \in \mathcal{U}(\mathfrak{g}) \otimes  \mathcal{U}(\mathfrak{g})$ deforms the $R$ matrix of the model under consideration as
\begin{equation}
\label{eq:quantumRmatrixTwist}
R_{12} \rightarrow F_{21} R_{12} F_{12}^{-1}.
\end{equation}
Twists depend on a deformation parameter -- we call it $\alpha$ to distinguish it from $\eta$ in the classical model -- and if we expand $F= 1\otimes 1 + \alpha F^{(1)} +\mathcal{O}(\alpha^2)$, the anti-symmetrization of its leading piece
\begin{equation}
r_{12} = F^{(1)}_{12} - F^{(1)}_{21} \in \mathfrak{g} \otimes \mathfrak{g},
\end{equation}
is guaranteed to solve the homogeneous classical Yang-Baxter equation
\begin{equation}
[r_{12},r_{13}]+[r_{12},r_{23}]+[r_{13},r_{23}]=0,
\end{equation}
where the subscripts indicate the tensor spaces in which an object acts nontrivially. This equation is related to our operator equation \eqref{eq:operatorCYBE}, and $r$ is related to $R$, via
\begin{equation}
\label{eq:rmatrixtoRoperator}
R(x) = \mbox{Tr}_2(r_{12} x_2),
\end{equation}
where antisymmetry of $r$ translates to antisymmetry of $R$ under the trace.

The Drinfeld twists that we will need, are associated to abelian and almost abelian $r$ matrices. An $r$ matrix is abelian if it is built out of commuting generators. The standard example is
\begin{equation}
r = h^1 \wedge h^2 \equiv h^1 \otimes h^2 - h^2 \otimes h^1,
\end{equation}
where $h^1,h^2$ are Cartan generators of $\mathfrak{g}$. We can take the Drinfeld twist associated to an abelian $r$ matrix to be
\begin{equation}
F = e^{i \alpha r}.
\end{equation}
Almost abelian $r$ matrices are sums of abelian pieces -- each an $r$ matrix itself -- where each added piece is constructed out of symmetries of the sum of the previous pieces, but not everything commutes. For example, consider $r = \hat{r} + \bar{r}$ with $\hat{r} = a\wedge b$ and $\bar{r} = c\wedge d$, and $[a,b]=0$ and $[c,d]=0$. This sum solves the classical Yang-Baxter equation provided $\bar{r}$ is subordinate to $\hat{r}$, meaning
\begin{equation}
[\mbox{ad}_c \otimes 1 + 1 \otimes \mbox{ad}_c, \hat{r}] = [\mbox{ad}_d \otimes 1 + 1 \otimes \mbox{ad}_d, \hat{r}] = 0,
\end{equation}
or equivalently
\begin{equation}
\label{eq:rbarsubordinatetorhat}
[\bar{r}_{13} + \bar{r}_{23},\hat{r}_{12}] = 0.
\end{equation}
For example, if we consider the Poincar\'e algebra (a subalgebra of the simple algebra $\mathfrak{so}(2,4)$) and use light cone coordinates $x^\pm = x^0 \pm x^1$, the $r$ matrix
\begin{equation}
r = \hat{r} + \bar{r} = m_{+2} \wedge p_+  + p_2 \wedge p_-, \quad \quad [m_{+2},p_2] = p_+,
\end{equation}
is almost abelian. Further examples can be found in \cite{Borsato:2016ose}. As discussed in \cite{vanTongeren:2016eeb}, thanks to the almost abelian structure we can take the Drinfeld twist for such $r$ matrices to be
\begin{equation}
\label{eq:almostabelianDrinfeldtwist}
F = \bar{F} \hat{F}, \quad \bar{F} = e^{i\gamma \bar{r}}, \quad \hat{F} = e^{i \beta \hat{r}},
\end{equation}
where we introduced a second deformation parameter, replacing $\alpha \hat{r} + \alpha \bar{r}$ by $\beta \hat{r} + \gamma \bar{r}$. Below we will see that the twist operator $\mathcal{F}$ perfectly matches this structure, for both abelian and almost abelian deformations. The twist operator also shows up in the classical analogue of equation \eqref{eq:quantumRmatrixTwist}.

\subsection*{Twisted monodromy matrix}

We can use the twist operator $\mathcal{F}$ to relate the monodromy matrices of deformed models to the monodromy matrix of the undeformed model. We define the monodromy matrix $M^g$ as
\begin{equation}
M^g = \mathcal{P} \left\{ \exp \int_0^{2\pi} L^g_\sigma d \sigma^\prime \right\},
\end{equation}
where $L^g$ is gauge-equivalent to the Lax matrix above,
\begin{equation}
L^g = g L g^{-1} + d g g^{-1}.
\end{equation}
Doing a further gauge transformation by $\mathcal{F}^{-1}$ we get
\begin{equation}
M^{\tilde{g}} = \mathcal{P} \left\{ \exp \int_0^{2\pi} L^{\tilde{g}}_\sigma d \sigma^\prime \right\} = \mathcal{F}^{-1}(2 \pi) M^g \mathcal{F}(0),
\end{equation}
where we only indicate $\sigma$ arguments explicitly. Now, when expressed entirely in terms of $\tilde{g}$, $M^{\tilde{g}}$ is identical in form to the undeformed monodromy matrix, which we denote $M^g_0$. We hence have
\begin{equation}
\label{eq:generaltwistedmonodromy}
M^g = \mathcal{F}(2 \pi) \widetilde{M_0^{g}} \mathcal{F}^{-1}(0),
\end{equation}
where we recall that the tilde indicates replacing $g$ by $\tilde{g}$.
We see that the deformed monodromy matrix is obtained from the formally undeformed one by multiplying it by twist operators. This is reminiscent of equation \eqref{eq:quantumRmatrixTwist} for Drinfeld twists, as used in the argumentation of \cite{vanTongeren:2015uha}. For abelian deformations we will come back to this in detail.

While formally undeformed, we should keep in mind that $M^{\tilde{g}}$ is built on the non-local fields making up $\tilde{g}$. In particular, assuming the original $g$ to be periodic, $\tilde{g}$ has twisted periodicity
\begin{equation}
\tilde{g}(2\pi)\tilde{g}(0)^{-1} =  \mathcal{F}^{-1}(2\pi) \mathcal{F}(0)= \mbox{Ad}_{\mathcal{F}^{-1}(0)}\left(\mathcal{P}\left\{\exp\left(-\eta  \int_{S^1}R(I^{l\tau})  d\sigma^\prime\right)\right\}\right).
\end{equation}
Alternatively, $\tilde{g}(2\pi)^{-1}\tilde{g}(0) =  g^{-1}(2\pi)\mathcal{P}\left\{\exp (\eta  \int_{S^1} R(I^{l\tau})  d\sigma^\prime)\right\}g(0)$, in line with \cite{Vicedo:2015pna}.

\section{Global symmetries and conserved currents}
\label{sec:globalsymmetries}

Yang-Baxter deformations break varying amounts of symmetry, depending on the $R$ operator under consideration. In our case the right $G$ symmetry of the principal chiral model, $g \rightarrow g h$, where $h$ is a constant element of $G$, is manifestly preserved. The left $G$ symmetry, however, is broken to a subgroup with algebra spanned by the generators $t$ for which
\begin{equation}
\label{eq:Rsymmetries}
[t,R(x)] = R([t,x]),\quad \forall x \in \mathfrak{g}.
\end{equation}
What are the Noether currents corresponding to these symmetries? In terms of right symmetry, $I$ is clearly conserved, and is the deformed analogue of $A$. For the left symmetry, it is the components of $I^l$ corresponding to solutions of equation \eqref{eq:Rsymmetries} that are preserved. Indeed, using equation \eqref{eq:lefteom}, then cyclicity of the trace ($\mbox{Tr}(a[b,c]) = \mbox{Tr}([a,b]c)$), and finally equation \eqref{eq:Rsymmetries} and antisymmetry of $\epsilon^{\alpha \beta}$, we have
\begin{equation}
\label{eq:Ilcomponentconservation}
\begin{aligned}
\partial_\alpha \mbox{Tr}(t I^{l\alpha}) &  = \eta \epsilon^{\alpha \beta}\mbox{Tr}(t [R(I^{l}_\alpha),I^{l}_\beta])\\
& = \eta \epsilon^{\alpha \beta} \mbox{Tr}([t,R(I^l_\alpha)]I^l_\beta + I^l_\alpha R([t,I^l_\beta])) = 0.
\end{aligned}
\end{equation}
Note that it is the partially conserved current $I^l$ that appears in $\mathcal{F}$.

In terms of $\tilde{g}$, the equations of motion actually imply that all of $\tilde{A}^l$ is conserved, seeming to suggest that the model has more symmetries than just claimed. However, the non-local definition of $\tilde{g}$ interferes with boundary conditions, so that not all of the would-be charges are actually conserved. Concretely we have
\begin{equation}
\label{eq:relationbetweenIleftandAtildeleft}
I^l = g \tilde{g}^{-1} \tilde{A}^l \tilde{g} g^{-1} = \mathcal{F} \tilde{A}^l \mathcal{F}^{-1},
\end{equation}
illustrating the twisted nature of the conserved current. The relevant components are equal when
\begin{equation}
\mbox{Tr}( \mathcal{F} t \mathcal{F}^{-1} I^l) = \mbox{Tr}(t I^l),
\end{equation}
i.e. in particular when $\mathcal{F}$ commutes with the relevant $t$.

For abelian deformations, the above allows us to express everything we need about $\mathcal{F}$ via conserved charges of the deformed model under consideration.

\section{Abelian deformations}

\label{sec:abeliandeformations}

Abelian deformations correspond to $R$ operators built out of commuting generators. These deformations are equivalent to performing sequences of TsT transformations \cite{Osten:2016dvf}. It is well known that TsT transformations of sigma models can be accounted for purely in terms of boundary conditions \cite{Frolov:2005dj,Alday:2005ww}. Here we will derive this result in the general language of twist operators.

The definition of $\mathcal{F}$ now involves an abelian current, so that we can write
\begin{equation}
\mathcal{F}(\tau,\sigma) = \exp\left(-\eta \int_\gamma \epsilon_{\alpha \beta} R(I^{l\beta}) \frac{d\sigma^{\prime\alpha}}{ds} d s \right)\,\mathcal{F}(0,0),
\end{equation}
where $\gamma$ is any path parametrized by $s$ that starts at $(0,0)$ and finishes at $(\tau,\sigma)$, and $\mathcal{F}(0,0)$ is just a global left $G$ transformation of our parametrization $\tilde{g}$. Moreover, $R$ now projects $I^{l}$ onto (some of) its conserved components, so that
\begin{equation}
\mathcal{F}(\tau,2\pi)\mathcal{F}^{-1}(\tau,0) = \exp\left(\eta \int_{S^1} R(I^{l\tau})  d\sigma^\prime\right) = e^{\eta R(Q^l)},
\end{equation}
where $Q^l$ denotes the matrix of conserved charges associated to $I^l$, i.e. the conserved quantities in the deformed model. We explicitly see that the twist operator behaves as the classical analogue of an abelian Drinfeld twist $e^{i \alpha r}$; $R(Q^l)$ acts in the tensor product of matrices and fields, in the latter case via the Poisson bracket, so that upon quantization we match the $i$ in the Drinfeld twist.\footnote{By equations \eqref{eq:quantumRmatrixTwist} and \eqref{eq:generaltwistedmonodromy}, $e^{\eta R(Q^l)}$ is the analogue of $e^{-i 2\alpha r}$. For our present purposes we can ignore the constant of proportionality relating $\alpha$ and $\eta$ -- it depends on the definition of $Q$ relative to the generators, and in particular includes a factor of the effective string tension.} Finally, since everything commutes, the components of $I^l$ corresponding to generators appearing in $R$ are conserved, and agree with the same components of the formally undeformed current $\tilde{A}^l$.

\paragraph{Twisted boundary conditions.} Denoting the commuting generators appearing in the $R$ operator by $h^i$, we can always parametrize our group elements in the form
\begin{equation}
\label{eq:abelianparametrization}
g(X,Y) = e^{X_i h^i} \bar{g}(Y), \quad \tilde{g}(\tilde{X},\tilde{Y}) = e^{\tilde{X}_i h^i} \bar{g}(\tilde{Y}),
\end{equation}
where $\tilde{g}$ of course only appears on shell. The relation $\tilde{g} = \mathcal{F}^{-1} g$, now becomes
\begin{equation}
e^{\tilde{X}_i h^i} \bar{g}(\tilde{Y}) = e^{(X_i + \eta \int_\gamma \epsilon_{\alpha \beta} R_i(I^{l\beta}) \frac{d\sigma^{\prime\alpha}}{ds} d s) h^i} \bar{g}(Y),
\end{equation}
where we write the $R$ operator in terms of its components, $R = R_i h^i$. In other words, $g$ and $\tilde{g}$ are related by the non-local field redefinition
\begin{equation*}
\tilde{X}_i (\tau,\sigma) = X_i (\tau,\sigma) + \eta \int_\gamma \epsilon_{\alpha \beta} R_i(I^{l\beta}) \frac{d\sigma^{\prime\alpha}}{ds} d s, \quad \tilde{Y} = Y.
\end{equation*}
When $(X,Y)$ satisfy the deformed equations of motion, $(\tilde{X},Y)$ satisfy the undeformed ones, up to a modification of boundary conditions. Assuming $X$ to be periodic, we find that
\begin{equation}
\label{eq:Xperiodicity}
\tilde{X}_i(\tau,0) - \tilde{X}_i(\tau,2\pi) = \eta \int_{S^1} R_i(I^{l\tau}) d\sigma^\prime = \eta R_i(Q^l).
\end{equation}
When a field $X$ is periodic from the target space perspective, i.e. $X \sim X + L$, worldsheet periodicity also allows field configurations that wind around the corresponding cycle, i.e. $X(\tau,0) - X(\tau,2 \pi) =  k L,\,\, k \in \mathbb{Z}$. This winding directly translates through the above analysis.

As an example we can consider $G = \mathrm{SO}(6)$ (or $G/H = \mathrm{S}^5$) with Cartan generators $h^i$, $i=1,2,3$, where we denote the associated coordinates and charges by $\phi_i$ and $J^i$ respectively. Then the abelian deformation for $r = \epsilon_{ijk} \eta^{-1} \gamma^i h^j \wedge h^k$ corresponds to $\tilde{\phi}_i(\tau,2\pi) - \tilde{\phi}_i(\tau,0) = \epsilon_{ijk} \gamma^j J^k$, in agreement with \cite{Frolov:2005dj}. In summary, abelian deformations can be purely accounted for in terms of the boundary conditions of the $X_i$ fields, those boundary conditions determined by the $R$ operator and the conserved charges of the deformed model.

\paragraph{Twisted monodromy matrix.} The above boundary conditions completely characterize the monodromy matrix as well. Since the non-local redefinition affects only the $X_i$ fields in the parametrization \eqref{eq:abelianparametrization}, following \cite{Alday:2005ww} we would like to explicitly factor these out of the monodromy matrix of the model. We start from the undeformed Lax connection
\begin{equation}
L_0 = L_{0+} + L_{0-}, \quad L_{0\pm} = \frac{1}{1\pm z} A_\pm,
\end{equation}
and perform a double gauge transformation
\begin{equation}
L_0 \rightarrow L_0^g = g L_0 g^{-1} + dg g^{-1} \rightarrow L_0^h = m^{-1}L_0^g m - m^{-1} d m,
\end{equation}
where $m = e^{X_i h^i}$ is the piece of $g$ associated to the generators in the $R$ operator and twist operator. This gives
\begin{equation}
L_{0\pm }^h = \frac{\mp z}{1\pm z} m^{-1} A^l_\pm m -  \frac{1}{1\pm z} d_\pm m m^{-1} = \frac{\mp z}{1\pm z} d_\pm \bar{g}(Y) \bar{g}^{-1}(Y) +  \frac{1}{1\pm z}  d_\pm X_i h^i,
\end{equation}
where $d_\pm X = P_{\pm\alpha}{}^\beta\partial_\beta X d\sigma^\alpha$. This Lax connection depends only on the $dX_i$, and hence the deformed Lax connection expressed via $\tilde{g}$, $\widetilde{L^h_0}$, depends only on the $d \tilde{X}_i$. The above field redefinition then gives a ``local'' substitution rule $dX \rightarrow d\tilde{X}(dX)$ that produces the deformed Lax pair and monodromy matrix,
\begin{equation}
M^h = \mathcal{P} \left\{ \exp \int_{0}^{2\pi}  L^h_\sigma d\sigma \right\},
\end{equation}
from their undeformed counterparts \cite{Alday:2005ww}.

To really put everything about the deformation of $M^g$ in terms of boundary conditions, we compare $M_0^g$ to $M_0^h$. As they are related by a gauge transformation,
\begin{equation}
M_0^g = m(2\pi) M_0^h m^{-1}(0),
\end{equation}
where, again, $M_0^h$ depends only on the $dX_i$. This means that at the deformed level, up to a similarity transformation that does not affect the generated conserved quantities,
\begin{equation}
\label{eq:abeliantwistedmonodromy}
M^{\tilde{g}} = \widetilde{M_0^{g}} \simeq \tilde{m}^{-1}(0)\tilde{m}(2\pi) \widetilde{M_0^{h}} =  e^{-\eta R(Q^l)}  \widetilde{M_0^{h}},
\end{equation}
where on the right-hand side $\widetilde{M_0^{h}}$ is identical to the undeformed monodromy matrix $M^h_0$, up to only the boundary conditions imposed on the $\tilde{X}_i$. This of course matches the general picture of equation \eqref{eq:generaltwistedmonodromy}, except that we were able to push ahead and express $M^{\tilde{g}}$ entirely in terms of undeformed quantities, up to only simple boundary conditions in field and matrix space, determined by (the classical analogue of) the Drinfeld twist.

Strictly speaking this is a circular definition of abelian deformed models, using twisted boundary conditions in terms of conserved charges that should be computed in the deformed models. However, we can split any model into sectors with fixed values for its conserved charges, and study each such sector independently. Moreover, we can supply the values that the conserved charges take, as separate external input. This breaks us out of the circle.

\section{Non-abelian deformations}

\label{sec:non-abeliandeformations}

If we think about non-abelian deformations in completely general terms, we immediately encounter several complications. First, the components of $I^l$ appearing in the twist operator are now no longer guaranteed to be associated to local conserved Noether currents of the deformed model. Second, there need not be a group parametrization that allows us to absorb the twist operator in a redefinition of the sigma model fields, with associated twisted boundary conditions. Of course, the algebraic constraints imposed by the classical Yang-Baxter equation do implicitly restrict the non-abelian structures under consideration -- they should be quasi-Frobenius -- so that perhaps some of these complications can be avoided in particular cases. To illustrate these points, let us consider almost abelian deformations.

\subsection*{Almost abelian deformations}

Almost abelian deformations correspond to particular ordered sequences of TsT transformations \cite{Borsato:2016ose,vanTongeren:2016eeb}. After doing a TsT transformation there may be commuting isometries left that allow for another TsT transformation, while the reverse order need not be possible, matching the algebraic discussion of almost abelian $r$ matrices in section \ref{sec:Drinfeldtwistedsymmetryandrmatrices}. This structure allows us to factorize the twist operator $\mathcal{F}$, analogously to the Drinfeld twist \eqref{eq:almostabelianDrinfeldtwist}.

For clarity we introduce a second deformation parameter, replacing $\eta \hat{R} + \eta \bar{R}$ by $\mu \hat{R} + \nu \bar{R}$. Consider $\hat{\mathcal{F}}$ as the solution to the fundamental linear problem
\begin{equation}
d \hat{\mathcal{F}} = -\mu\hat{R}(\star Y) \hat{\mathcal{F}},
\end{equation}
where $Y$ is some to be determined current, required to reduce to the current $\hat{I}^l$ for the $\hat{R}$ deformation in the limit $\nu \rightarrow 0$. We would like to perform a ``subsequent'' deformation corresponding to $\bar{R}$, where we expect the argument of $\bar{R}$ to be simply $\star I^l$,
\begin{equation}
d \bar{\mathcal{F}} = -\nu \bar{R}(\star I^{l}) \bar{\mathcal{F}}.
\end{equation}
With these relations, we get that $\mathcal{F} = \bar{\mathcal{F}}\hat{\mathcal{F}}$ satisfies
\begin{equation}
d \mathcal{F} = -\left(\nu  \bar{R}(\star I^{l}) + \mu \bar{\mathcal{F}} \hat{R}(\star Y) \bar{\mathcal{F}}^{-1}\right) \mathcal{F}.
\end{equation}
This solves the fundamental linear problem for $\mathcal{F}$, provided
\begin{equation}
\label{eq:almostabelianRandYrequirement}
\bar{\mathcal{F}} \hat{R}(Y) \bar{\mathcal{F}}^{-1} = \hat{R}(I^{l}).
\end{equation}
Translating equation \eqref{eq:rbarsubordinatetorhat} to operator form gives
\begin{equation}
[\bar{R}(x),\hat{R}(y)] = \hat{R}([\bar{R}(x),y]), \quad \forall x,y \in \mathfrak{g}
\end{equation}
so that
\begin{equation}
\bar{\mathcal{F}}\hat{R}(z) \bar{\mathcal{F}}^{-1} = \hat{R}(\bar{\mathcal{F}} z \bar{\mathcal{F}}^{-1}), \quad \forall z \in \mathfrak{g}.
\end{equation}
Hence we should take $Y= \bar{\mathcal{F}}^{-1} I^{l} \bar{\mathcal{F}}$. As required, $Y$ reduces to $\hat{I}^l$ in the limit $\nu \rightarrow 0$.

The above is perfectly in line with the factorized structure of the TsT transformations, and shows explicitly that the twist operator is the analogue of an almost abelian Drinfeld twist.\footnote{Coincidentally we have now covered all unimodular cases \cite{Borsato:2016ose} for which Drinfeld twists are explicitly known \cite{vanTongeren:2016eeb}.} We can tentatively think of the $\bar{\mathcal{F}}$ deformation of the current appearing in $\hat{R}$ as the counterpart to the nontrivial field redefinitions required to perform the second TsT transformation in geometric terms.\footnote{If no such field redefinition is required, the TsT transformations by definition commute, reducing us to the abelian case where $k^l = A^l  - d \hat{\mathcal{F}} \hat{\mathcal{F}}^{-1} - d \bar{\mathcal{F}} \bar{\mathcal{F}}^{-1}$, and $R(\mathcal{F} X \mathcal{F}^{-1}) = R(X)$ (for any combination of hats and bars).} In fact, while in the general form of $\mathcal{F}$ we encounter $\hat{R}(I^l)$ which is manifestly not conserved, it is nice to see that $\hat{R}(Y)$ is conserved, since by equation \eqref{eq:relationbetweenIleftandAtildeleft},
\begin{equation}
\hat{R}(Y) = \hat{R}(\bar{\mathcal{F}}^{-1} I^{l} \bar{\mathcal{F}}) = \hat{R}(\hat{\mathcal{F}}^{-1} \tilde{A}^{l} \hat{\mathcal{F}}) = \hat{R}(\tilde{A}^{l}),
\end{equation}
and $\partial_\alpha \tilde{A}^{l\alpha}=0$ by definition; $\hat{\mathcal{F}}$ disappears in the third equality since it commutes with the generators making up $\hat{R}$, cf. equation \eqref{eq:rmatrixtoRoperator}. Unfortunately, $Y$ is not a local current, and does not integrate to a standard conserved charge. We see that despite some nice features, the first complication indicated above remains in the almost abelian case. We can definitely formulate the $\hat{r}$ deformation in terms of twisted boundary conditions, but the charges in these boundary conditions lose their meaning after the subsequent $\bar{r}$ deformation.

If we gloss over the first complication and try to push the analogy to the abelian case further and consider boundary conditions, we encounter our second complication. In the abelian case we have a natural group parametrization that allows us to absorb the twist operator in the sigma model fields in a ``diagonalized'' fashion. By definition this exact thing is impossible for non-abelian deformations -- in the $\hat{r}$ deformed model we can do it for $\hat{F}$; in the $\hat{r}+\bar{r}$ deformed geometry we can do it for $\bar{F}$, but not for $\hat{F}$. Presumably we should instead consider substantially different types of group parametrization and field redefinition.

The upshot is that at this stage the deformation cannot be directly absorbed in simple boundary conditions for the sigma model fields. As such, also the formally undeformed monodromy matrix $\widetilde{M^{h}_0}$ is not simply obtained from the true undeformed $M_0$ by just putting twisted boundary conditions for the sigma model fields.

\section{Conclusions}

In this note we discussed the twist operator formulation of homogeneous Yang-Baxter deformed sigma models, with the larger aim of systematically tackling these models at the quantum level. We manifestly demonstrated that the twist operator looks like the classical analogue of a Drinfeld twist, in all abelian and almost abelian cases. While the twist operator approach formally accomplishes a mapping from deformed models to the undeformed model, its use in e.g. the spectral problem is not immediately clear beyond abelian deformations, because there the twist operator cannot be straightforwardly diagonalized in terms of sigma model fields and conserved charges. What is the most useful approach to take in these cases? Is this even a sensible question to ask in general? It would be interesting to develop a more sophisticated perspective on this problem, perhaps starting by investigating various specific $r$ matrix examples, and deformations of simple models like the flat space sigma model, where some of these obstacles may be partially avoided. It may also help to consider these models from the perspective of non-abelian T duality.

In terms of a broader outlook, if we think about the quantum $\ads$ string, obstacles actually already arise in certain abelian cases. The $r$ matrix and associated twist of an abelian deformation need not be diagonalizable, as e.g. discussed for a particular dipole deformation in \cite{Guica:2017mtd}. In this case, the Bethe ansatz is not applicable due to a lack of suitable vacuum. In principle one may overcome such problems by for example considering Baxter's approach instead, as indeed was done at one loop in \cite{Guica:2017mtd}. However, the true power of integrability for the $\ads$ string lies in using integrability not just perturbatively, but actually at finite coupling. At least to date, this approach relies essentially on the exact S matrix of the string, which is fixed by the global symmetries of the string in the BMN light-cone gauge, see e.g. \cite{Arutyunov:2009ga}. For deformations that break these symmetries -- any non-diagonalizable one does -- we need to re-determine the finite coupling data that goes in to the integrability machinery, be it Baxter or Bethe. Put differently, it is not clear how to apply methods that label states by Cartan charges, when these charges are no longer meaningful. It would be interesting to understand how to efficiently describe generic deformed models at the quantum level. We may hope that there is a yet undiscovered, refined description of the undeformed $\ads$ string, where arbitrary Drinfeld twists can just be inserted appropriately, analogously to how this can currently be done for the diagonal case.

\section*{Acknowledgments}

I would like to thank R. Borsato, and D. Osten for discussions, and R. Borsato for comments on the draft. I am supported by L.T.

\appendix

\section{Symmetric space models}
\label{app:symmetricspacediscussion}

The relevant added structure of a symmetric space $G/H$ compared to the group manifold case of the main text, is a decomposition of $\mathfrak{g}$ as
\begin{equation}
\mathfrak{g} = \mathfrak{h} + \mathfrak{k}, \quad \mbox{with} \quad [\mathfrak{h},\mathfrak{h}] \subset \mathfrak{h}, \quad [\mathfrak{h},\mathfrak{k}] \subset \mathfrak{k}, \quad [\mathfrak{k},\mathfrak{k}] \subset \mathfrak{h},
\end{equation}
where $\mathfrak{h}$ is the Lie algebra of $H$. We introduce $P$ as the projection operator from $\mathfrak{g}$ to $\mathfrak{k}$. This structure is compatible with the Killing form, namely $\mbox{Tr}(x P(y)) = \mbox{Tr}(P(x)P(y))$, $\forall x,y \in \mathfrak{g}$.

The Yang-Baxter deformation of the symmetric space sigma model \cite{Delduc:2013fga} has Lagrangian
\begin{equation}
\mathcal{L} = P_-^{\alpha\beta}\mbox{Tr}\left(A_\alpha P \frac{1}{1-\eta R_g\circ P }(A_\beta)\right).
\end{equation}
Again we consider only homogeneous deformations \cite{Kawaguchi:2014qwa,Matsumoto:2015jja}. We now introduce
\begin{equation}
I = \frac{1}{1+\eta R_g\circ P }(A_+) + \frac{1}{1-\eta R_g\circ P}(A_-).
\end{equation}
The equations of motion of the model are then
\begin{equation}
\mathcal{E} = \partial_\alpha P(I^\alpha) - \left[I_\alpha,P(I^\alpha)\right] = 0.
\end{equation}
The flatness of $A$ expressed in terms of $I$, as in the group case, becomes
\begin{equation}
\epsilon^{\alpha\beta}\left(\partial_\alpha I_\beta - \partial_\beta I_\alpha - [I_\alpha,I_\beta]\right)= - 2 \eta R_g(\mathcal{E}),
\end{equation}
and we can take our Lax connection to be
\begin{equation}
L(z) = I + (z-1) P(I_+) + (z^{-1}-1) P(I_-).
\end{equation}
We can introduce a twist operator $\mathcal{F}$ \cite{Matsumoto:2015jja}, as in the main text. We just get a slight modification in the fundamental linear problem,
\begin{equation}
d\mathcal{F} = (I^l - A^l) \mathcal{F} = -\eta \,R(gP(\star I)g^{-1})\mathcal{F}.
\end{equation}
This was to be expected, since now $k = g P(I) g^{-1}$ should be the partially conserved deformed current. Indeed, on shell,
\begin{equation}
\begin{aligned}
\partial_\alpha k^\alpha & = \mbox{Ad}_g\left(\partial_\alpha P(I^\alpha) - [A_\alpha, P(I^\alpha)]\right)\\
& = \mbox{Ad}_g\left([I_\alpha - A_\alpha, P(I^\alpha)]\right)\\
& = \eta \epsilon_{\alpha \beta} \mbox{Ad}_g \left([R_g(I^\alpha),P(I^\beta)]\right)\\
& =  \eta \epsilon_{\alpha \beta} [R(k^\alpha),k^\beta],
\end{aligned}
\end{equation}
which is of the form of the equation of motion \eqref{eq:lefteom} for $I^l$ in the group case. Hence, as in the group case, $\partial_\alpha \mbox{Tr}(t k^\alpha) =0$ for those $t$ that generate symmetries, cf. equation \eqref{eq:Rsymmetries}.

The minimal differences from the group case do not change the twist picture at the level of the monodromy matrix or general boundary conditions. The analysis of the sigma model field boundary conditions for the abelian case also goes through as before. Only the discussion of the abelian deformed monodromy matrix requires some extra structure, due to the projection operators. Following \cite{Frolov:2005dj} we effectively rephrase the symmetric space sigma model as a principal chiral model, see e.g. section 1.5.2 of \cite{Arutyunov:2009ga}. Focussing on $\ads$, working in the conventions of \cite{Arutyunov:2009ga}, we introduce $G = g K g^t$ which depends only on the coset degrees of freedom. After a gauge transformation by $g$, the undeformed symmetric space Lax pair can then be brought to exactly the form of the group case
\begin{equation}
L^g = \frac{x}{1- x}(-d_- G G^{-1})-\frac{x}{1+ x}(-d_+ G G^{-1}),
\end{equation}
where $x=(1+z)/(1-z)$. If we then parametrize $g$ as $g(X,Y) = e^{X_i h^i} \bar{g}(Y) = m \bar{g}$, with $G = m \bar{G} m^t$, the dependence of $m^{-1} L^g m - m^{-1} dm$ on the $X_i$ is only through their derivatives, and we end up exactly with the relation of equation \eqref{eq:abeliantwistedmonodromy}. This abelian analysis can be directly adapted to include fermions \cite{Alday:2005ww}.

Of course, the subtleties surrounding non-abelian deformations discussed in the main text, equally apply to the symmetric space case.


\bibliographystyle{nb}

\bibliography{TwistfunctionsandBCs_arxiv_v3_actually_to_arxiv.bbl}


\end{document}